\documentclass{emulateapj}

\usepackage{amsmath}
\usepackage{amssymb}
\usepackage{graphicx}
\usepackage{grffile}
\usepackage[usenames,dvipsnames]{color}
\usepackage{bm}
\usepackage{natbib}
\usepackage{hyperref}
\usepackage{balance}
\usepackage{tabularx}

\usepackage{array}
\newcolumntype{L}{>{\raggedright\let\newline\\\arraybackslash\hspace{0pt}}X}

\usepackage[english]{babel}

\newcommand{\pow}[1]{\ifmmode{}^{#1}\else ${}^{#1}$\fi}
\newcommand{\HI}{{\text{H\MakeUppercase{\romannumeral 1}}} }

\newcommand{\Lya}{\ifmmode{{\rm Ly}\alpha}\else Ly$\alpha$\ \fi}
\newcommand{\cm}{\ifmmode{{\rm cm}}\else cm\fi}
\newcommand{\ccm}{\,\mathrm{cm}^{-3}} 
\newcommand{\ergps}{\,{\rm erg}\,{\rm s}\ifmmode{}^{-1}\else ${}^{-1}$\fi}
\newcommand{\Mpch}{\,{\rm Mpc}\,\ifmmode h^{-1}\else $h^{-1}$\fi}
\newcommand{\kms}{\,\mathrm{km}\,\mathrm{s}^{-1}}

\newcommand{\vek}[1]{\bm{#1}}
\newcommand{\sbu}{\,{\rm erg}\,{\rm s}\pow{-1}\,{\rm cm}\pow{-2}\,{\rm arcsec}\pow{-2}}

\slugcomment{Accepted for publication in \apj\ 2016 December 16}

\begin{document}
\title{Giant Lyman-Alpha Nebulae in the Illustris Simulation}

\author{Max Gronke}
\affil{Institute of Theoretical Astrophysics, University of Oslo, Postboks 1029 Blindern, 0315 Oslo, Norway}
\email{maxbg@astro.uio.no}

\author{Simeon Bird}
\affil{Department of Physics and Astronomy, Johns Hopkins University, 3400 N. Charles St., Baltimore, MD 21218, USA}


\begin{abstract}
Several `giant' Lyman-$\alpha$ (Ly$\alpha$) nebulae with extent $\gtrsim 300\,$kpc and observed Ly$\alpha$ luminosity of $\gtrsim 10^{44}\,{\rm erg}\,{\rm s}^{-1}\,{\rm cm}^{-2}\,{\rm arcsec}^{-2}$ have recently been detected, and it has been speculated that their presence hints at a substantial cold gas reservoir in small cool clumps not resolved in modern hydro-dynamical simulations.
We use the \texttt{Illustris} simulation to predict the Ly$\alpha$ emission emerging from large halos ($M > 10^{11.5}M_{\odot}$) at $z\sim 2$ and thus test this model. We consider both AGN and star driven ionization, and compare the simulated surface brightness maps, profiles and Ly$\alpha$ spectra to a model where most gas is clumped below the simulation resolution scale. We find that with \texttt{Illustris} no additional clumping is necessary to explain the extents, luminosities and surface brightness profiles of the `giant Ly$\alpha$ nebulae' observed. Furthermore, the maximal extents of the objects show a wide spread for a given luminosity and do not correlate significantly with any halo properties. We also show how the detected size depends strongly on the employed surface brightness cutoff, and predict that further such objects will be found in the near future.
\end{abstract}


\section{Introduction}
\label{sec:intro}
The medium immediately surrounding galaxies -- often dubbed the circumgalactic-medium (CGM) -- provides a gas reservoir for star formation and as such is crucial for the study of galaxy formation and evolution. Extended faint \Lya emission originating from these regions directly probes this gas, uniquely so at higher redshifts where the observation of other emission lines is challenging \citep[for reviews see, e.g.,][]{Barnes2014,Dijkstra2014_review,Hayes2015}. These `\Lya nebulae' are often dubbed `Lyman-$\alpha$ blobs' (LABs), or when fainter and surrounding a galaxy also called `Lyman-$\alpha$ halos' (LAHs). While only a few LABs have been found so far \citep{1999MNRAS.305..849F,2000ApJ...532..170S,Matsuda2011,Ao2015}, it has been shown that LAHs surround Lyman-break or `drop-out' galaxies (LBGs), galaxies selected through their \Lya emission (\Lya emitters or LAEs) as well as H$\alpha$ selected galaxies \citep[recently shown by][]{2016MNRAS.458..449M} leading to the conjecture that most (if not all) star-forming galaxies are associated with a LAH. Due to their very low surface-brightness (SB) profiles a stacking technique is often used in order to study the properties of LAHs. 
\citet{Steidel2010a} stacked a sample of $92$ LBGs at $z\sim 2.65$ to reach SB limits of $\sim 10^{19}\sbu$ and found the LAHs extend out to $\sim 80\,{\rm kpc}$ with a exponential scale length of $r_\alpha \approx 20-30\,{\rm kpc}$ which is a factor of a few larger than the scale length of the continuum emission. This result is consistent with the more recent finding of \citet{Matsuda2012} who used a set of $2128$ LAEs and $24$ LBGs at $z\sim 3.1$ to find scale lengths of $r_\alpha \sim 10-30\,$kpc and $r_\alpha\sim 20\,$kpc, respectively. Other studies \citep{Rauch2008,Momose2014} support this picture but suggest a smaller LAH for LAEs as opposed to LBGs. This could however be an environmental effect if LBGs reside in higher-density environments \citep{Matsuda2012}. Furthermore, ultradeep MUSE observations recently revealed the LAHs surrounding $21$ individual galaxies~\citep{Wisotzki2015}. Without the need of stacking their SB profiles, they confirmed that the scale of the \Lya emitting region is several times larger than the scale of the continuum emission.

While the \Lya nebulae surrounding `normal' galaxies extend $\sim$ tens of kpc and are relatively faint $L_\alpha \sim 10^{42} \ergps$, LAHs around a rare $z\gtrsim 2$ population of galaxies sometimes dubbed `high redshift radio galaxies' (HzRGs, which are associated with one or multiple quasars) have extensions up to hundreds of kpc and luminosities of $L_\alpha\sim 10^{44}\ergps$ \citep{2003ApJ...592..755R,2007MNRAS.378..416V,2012ApJ...752...39T,2015Natur.524..192M,Hennawi2015,Borisova2016}. \citet{Cantalupo2014} detected the most massive of these nebulae (sometimes called the `slug nebula'), measuring a maximum extent of $\sim 450\,$kpc and $L_\alpha\sim 10^{45}\ergps$ ($L_\alpha\sim 2\times 10^{44}\ergps$ excluding the emission directly from the quasar).

Although the existence of LABs, LAHs and the giant LAHs is observationally well-established, and some aspects are theoretically understood \citep[e.g., their existence implies a fairly large hydrogen mass around dark-matter halos,][]{Matsuda2012}, some key questions remain unclear. One which is debated in the literature is the energy source of these --  bright and extended -- \Lya nebulae. Commonly, two possibilities are discussed: \textit{(i)} \Lya production in the central region and subsequent scattering in the surrounding gas leading to the observed halo \citep[e.g.][]{2007ApJ...657L..69L,Kramer2012}, or, \textit{(ii)} \textit{in situ} \Lya production in an extended region. The latter would be possible, via e.g.~cooling luminosity of gas falling into the potential well of the galaxy \citep[e.g.][]{Haiman2000ApJ...537L...5H,Dijkstra2009a}. Alternatively, ionizing photons escaping from the central region or originating from nearby galaxies could lead to recombination events in the surrounding medium \citep[e.g.][]{2001ApJ...556...87H,2005ApJ...622....7F,Mas-Ribas2016}.

In particular, the discovery of the giant LAHs poses the question of how \Lya emission can be located so far from an ionizing source. \citet{Cantalupo2014} carried out radiative transfer simulations on a zoom-in hydro-dynamical simulation and concluded that their model could not explain a \Lya object having this SB level and extent.  Instead they proposed that hydrodynamical simulations miss a substantial fraction of cold, dense clumps which will boost the \Lya luminosity. Specifically, they calculate the required clumping factor $C\equiv \langle n_e^2 \rangle / \langle n_e \rangle^2$ (where $n_e$ is the electron number density) to be between $20$ and $1000$ on scales below a few kpc. In this work we want to revisit this question using the more modern cosmological hydro-dynamical \texttt{Illustris} simulation \citep{2014Natur.509..177V,2015A&C....13...12N} which
features a fully realised and well-tested model for galaxy formation,
tuned to produce a realistic galaxy population at $z \sim 0$. This model includes efficient supernova feedback, metal cooling and, importantly for studying the gas around quasars, a recipe for AGN feedback.
Using this publicly available data allows us to study a statistically relevant ensemble of halos instead of focusing on individual objects.

The study is structured as follows. In Sec.~\ref{sec:method} we lay out our model and numerical methods. We present our results in Sec. \ref{sec:results}, and conclude in Sec.~\ref{sec:conclusions}.

\section{Methods}
\label{sec:method}

In this section, we first describe the hydrodynamical simulation used for our work (\S~\ref{sec:hydr-simul}), then introduce our simple ionization models (\S~\ref{sec:model}) before explaining the \Lya radiative transfer simulation employed (\S~\ref{sec:radi-transf-calc}).

\begin{table}
  \centering
  \caption{Overview of the model parameters}
  \begin{tabularx}{\linewidth}{llLl}
\hline\hline
    Model & Parameter & Description & Fiducial value \\
\hline \hline
AGN & $r_{\rm ion}$ & Radius of ionized region around AGN & $20\,$kpc \vspace{3pt} \\
Stars & $m_{\rm ion}$ & Cell volume ionized in units of Str\"omgren spheres (Eq.~\eqref{eq:ion_stars})& $1$ \vspace{3pt} \\
\hline
\hline
  \end{tabularx}
  \label{tab:model_params}
\end{table}

\subsection{The hydrodynamical simulation}
\label{sec:hydr-simul}

The \texttt{Illustris} simulation \citep{2014Natur.509..177V} is a hydro-dynamical cosmological simulation (simulation box side length of $106.5\,$cMpc) performed using the moving mesh code \texttt{AREPO} \citep{2010MNRAS.401..791S}. It includes gas cooling, and photo-ionization from a uniform ultra-violet background. Subgrid models are included for black holes \& black hole feedback, stochastic star formation (with a density threshold of $0.13\ccm$ above which an ad-hoc equation of state is imposed), stellar evolution, and stellar feedback, tuned to reproduce the galaxy stellar mass function at $z=0$. 
Note that molecular cooling is not included, and so the simulation does not accurately follow the temperature of gas with $T < 10^4\,$K.
In particular, we made use of the `Illustris-1' simulation which possesses a baryon (dark matter) particle mass of $12.6\times 10^{5}\,M_\sun$ ($62.6\times 10^{5}\,M_\sun$) and a resulting gravitational softening length of $\sim 700\,{\rm pc}$ ($1.4\,{\rm kpc}$). For the extraction of the data we made use of the excellent public interface provided by the \texttt{Illustris} team \citep{2015A&C....13...12N} which allowed us to post-process individual cutout halos.

Inspired by the observations of \citet{Cantalupo2014}, we selected a halo with total mass $M\sim 10^{12.5}\,M_\sun h^{-1}$ at $z\sim 2$. This halo possesses an active, central black hole with a mass inflow rate of $\dot M_{\rm BH}\sim 1.23 M_\sun {\rm yr}^{-1}$, a initial neutral fraction of $\sim 26\%$ and a cold ($T<10^5\,K$) gas fraction of $\sim 30\%$. The total gas mass of this halo -- as given by the simulation output -- is $\sim 4.5 \times 10^{11}\,M_\sun h^{-1}$. The relevant quantities (per particle) for our work are: the cell volume $V_{\rm cell}$, the neutral hydrogen number density $n_{\rm HI}$, the ionized hydrogen number density $n_{\rm HII}$, the electron density $n_e$, the gas velocity $\vek{v}$, the star-formation rate $SFR$, the metallicity $Z$, and the temperature $T$. We compute $T$ as
\begin{equation}
T =  \frac{2 u \mu}{3 k_B}\,,
\end{equation}
where $u$ is the internal energy, $\mu$ the mean molecular weight of the gas, and $k_B$ the Boltzmann constant. Dust plays a crucial role in \Lya radiative transfer and dust reddening is expected to be important in these massive halos. We follow \citet{Laursen2009} and calculate an effective number density of dust atoms as
\begin{equation}
n_d = (n_\HI + f_{\rm ion} n_{\rm HII})\frac{Z}{Z_\sun}
\label{eq:dust}
\end{equation}
which defines a dust optical depth through $\tau_d = n_d \sigma_d d$, where $\sigma_d$ is the dust optical depth and $d$ the path length considered. For the former we use the value of the Small Magellanic Cloud $\sigma_{\Lya, {\rm SMC}}\approx 1.58\times 10^{-21}\,{\rm cm}^{2}$ \citet{Pei1992} and take furthermore the dust-to-gas ratio in an ionized region to be characterized by $f_{\rm ion}=0.01$ \citep[see \S\,2.2.1 in][for a detailed discussion of why this choice is a good approximation]{Laursen2009}.
The effect of dust is a major source of uncertainty in our work; we discuss the effect of this uncertainty in \S~\ref{sec:surf-brightn} and \S\ref{sec:comp-observ}. In order to demonstrate its impact, we also show some radiative transfer results excluding dust reddening.

Self-shielding of gas from the metagalactic UV background is included in \texttt{Illustris} using the prescription of \citet{Rahmati2013}. This includes the neutral fraction in the (subgrid) star-forming gas, which is roughly unity, neglecting the local radiation field of the forming stars.

\begin{figure}
  \centering
  \includegraphics[width=.5\textwidth]{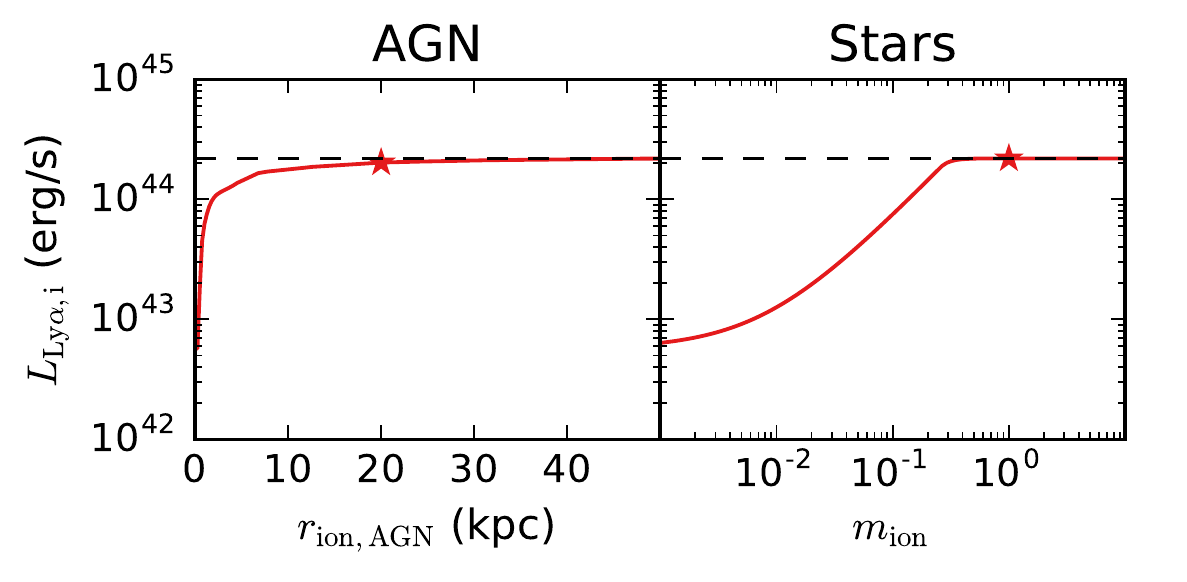}
  \caption{Intrinsic \Lya luminosities for the two models. The black-dashed line denotes the intrinsic luminosity reached if the entire grid is fully ionised and the red star symbols illustrate the choice of our fiducial models. From the two leftmost panels, it is clear that in each model reaching the maximal possible luminosity with the gas as given by the hydrodynamical simulation output is possible with either moderate ionization activity from either the star forming regions or the black hole.\\}
  \label{fig:Li}
\end{figure}

\begin{figure*}
  \centering
  \plotone{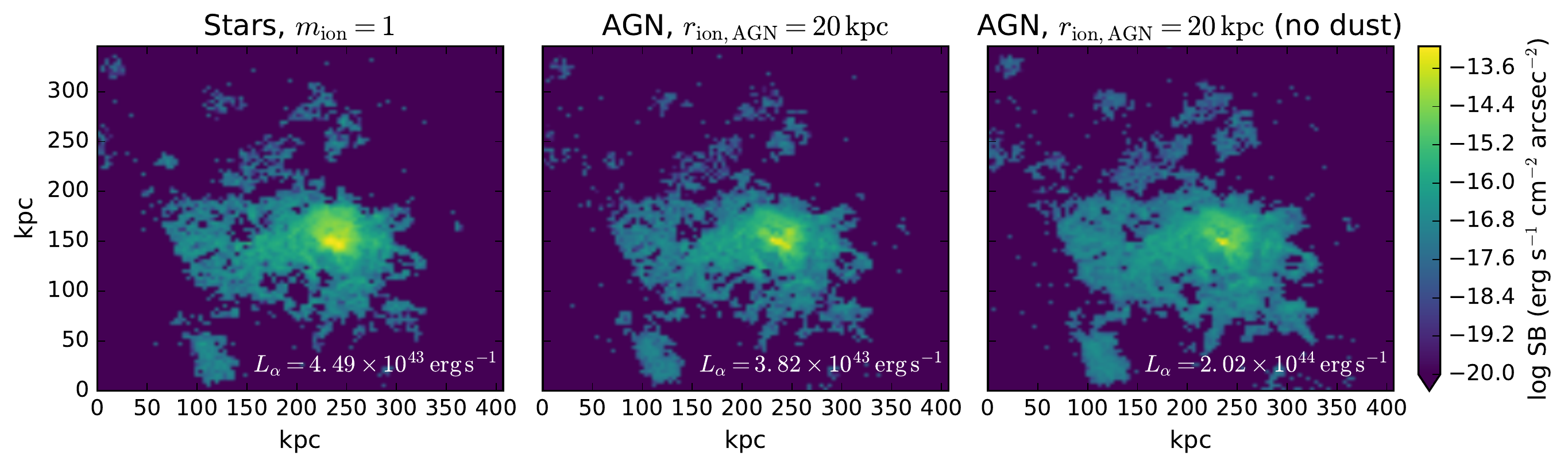}
  \caption{Surface brightness maps for the fiducial models (from left to right panel): star-driven ionization, AGN-driven ionization, and AGN-driven ionization without dust (see \S\ref{sec:model}).\\}
  \label{fig:SB}
\end{figure*}

\subsection{The model}
\label{sec:model}

In addition to the halo (as described in the previous section), we consider two possibilities for the radiation field:
\begin{enumerate}
\item \textit{AGN ionization.} In this scenario, we fully ionize all the cells falling within a spherical region around the central black hole with radius $r_{\rm ion}$. This clearly is a simplification of the true ionization mechanism by quasars as we neglect radiative transfer effects from different density structures as well as the likely existence of conical jets. However, in this work we are content with a first order approximation as \citet{Cantalupo2014} demonstrated that it was not possible to resolve the discrepancy between the observations and simulations without extra subgrid clumping even in the extreme cases when \textit{all} the hydrogen is ionized or neutral. Quasars can potentially ionize a region up to several Mpc \citep{2000ApJ...542L..75C}. However, we choose a relatively small value for $r_{\rm ion}$ in \S\ref{sec:surf-brightn} in order to \textit{(i)} show the radiative transfer effect in the outer region (i.e., to be much smaller than the total extent of the \Lya halo), and \textit{(ii)} still be large enough so that a small change in $r_{\rm ion}$ does not lead to a significant change in the intrinsic luminosity (\S~\ref{sec:intr-lya-lumin}).
As a fiducial value we pick $r_{\rm ion}= 20\,$kpc.
\item \textit{Ionization from stars.} Here, we ionize all the star-forming cells by subtracting
\begin{equation}
\Delta X_{\HI} =  \frac{Q_{\rm ion} m_{\rm ion}}{ n_{\rm HI}^2 \alpha_B(T) V_{\rm cell}}
\label{eq:ion_stars}
\end{equation}
from the cells' neutral hydrogen fraction, while ensuring $X_{\rm HI} \ge 0$.
In the above equation, $\alpha_B$ is the case-B recombination coefficient (see below) and $Q_{\rm ion} = 2\times 10^{53}\frac{{\rm SFR}}{1 M_{\sun} {\rm yr}}$ for a range of different stellar models \citep{Rahmati2013a}. Here, the free parameter $m_{\rm ion}$ can be interpreted as the number of (fully ionized) Str\"omgren spheres (volume $Q_{\rm ion} / (n_\HI^2 \alpha_B)$) placed in a neutral cell. This means for a cell with $X_\HI = 1$ if there is one star (cluster) per cell, the `sub-grid' escape fraction of ionizing photons is unity and if the ionized region does not overlap with the boundary then $m_{\rm ion}\sim 1$. In reality it is likely that the escape fraction is less than unity, and that nearby cells are also affected by ionizing photons. Furthermore, we expect several massive stars within each cell, given that the typical life-time of a massive star is $5\--10$\,Myr. However, given the uncertainty in calculating these effects, and the fact that, as shown in Fig.~\ref{fig:Li}, the \Lya luminosity saturates at $m_{\rm ion} \approx 0.2$, $m_{\rm ion} = 1$ appears to be a reasonable fiducial value and we adopt it for the rest of the paper.
\end{enumerate}
These model parameters are summarized in Table~\ref{tab:model_params}. The total amount of gas within the simulation is conserved in both the scenarios considered. We assume that the ionizing radiation fields are sufficiently intense that the gas will be highly ionized \citep{2013ApJ...766...58H}. In fact, we allow the hydrogen neutral fraction in the \textit{maximally} ionized regions we reach a minimum value of $X_{\rm HI,\, min} = 0$. As the emissivity $\propto (1 - X_{\rm HI})^2$ a slightly larger value of $X_{\rm HI}$ will not have a strong effect on the SB values (however, it can affect the \Lya radiative transfer, see \S~\ref{sec:lya-spectra}).

Given the number density of ionized hydrogen $n_{\rm HII}$, the electron number density $n_e$ and a temperature $T$ per cell, we compute the total \Lya luminosity assuming solely `case-B' recombination\footnote{`case-B' recombination denotes the recombination in a medium which is optically thick to ionizing photons. This leads to the immediate re-absorption of an emitted ionizing photon \citep[see, e.g.,][for details]{1989agna.book.....O,Dijkstra2014_review}.}
\begin{equation}
L_{\rm \alpha, i} = \sum\limits_{\rm cells} n_{\rm HII} n_e N_\alpha(T) \alpha_{B}(T) V_{\rm cell}\,,
\label{eq:Li}
\end{equation}
where $\alpha_{B}(T)$ is the `case-B' recombination coefficient from the fitting formulae by \citet{Hui1997}, and $N_\alpha(T)$ is the average number of \Lya photons produced per `case-B' recombination event. For $N_\alpha(T)$ we adopt the fit provided by \citet{Cantalupo2008}. As \Lya is produced mainly in high-density environments, the contribution of `case-A' recombination can be safely neglected \citep{1996ApJ...468..462G}. \Lya cooling radiation is highly sensitive to the temperature of the cold gas and no consensus has been reached on its contribution \citep{2005ApJ...622....7F,2010MNRAS.407..613G,Faucher-Giguere2010,Rosdahl2012,2013ApJ...775..112C}. In order to assess its impact a fully coupled radiation-hydrodynamical simulation is preferable \citep[as, e.g., in ][]{Rosdahl2012} which can in principle model the temperature state of the ISM. We do not address the impact of cooling radiation in this work, and, thus neglect its contribution. 
Both choices only reduce the final luminosity.

\begin{figure}
  \centering
  \plotone{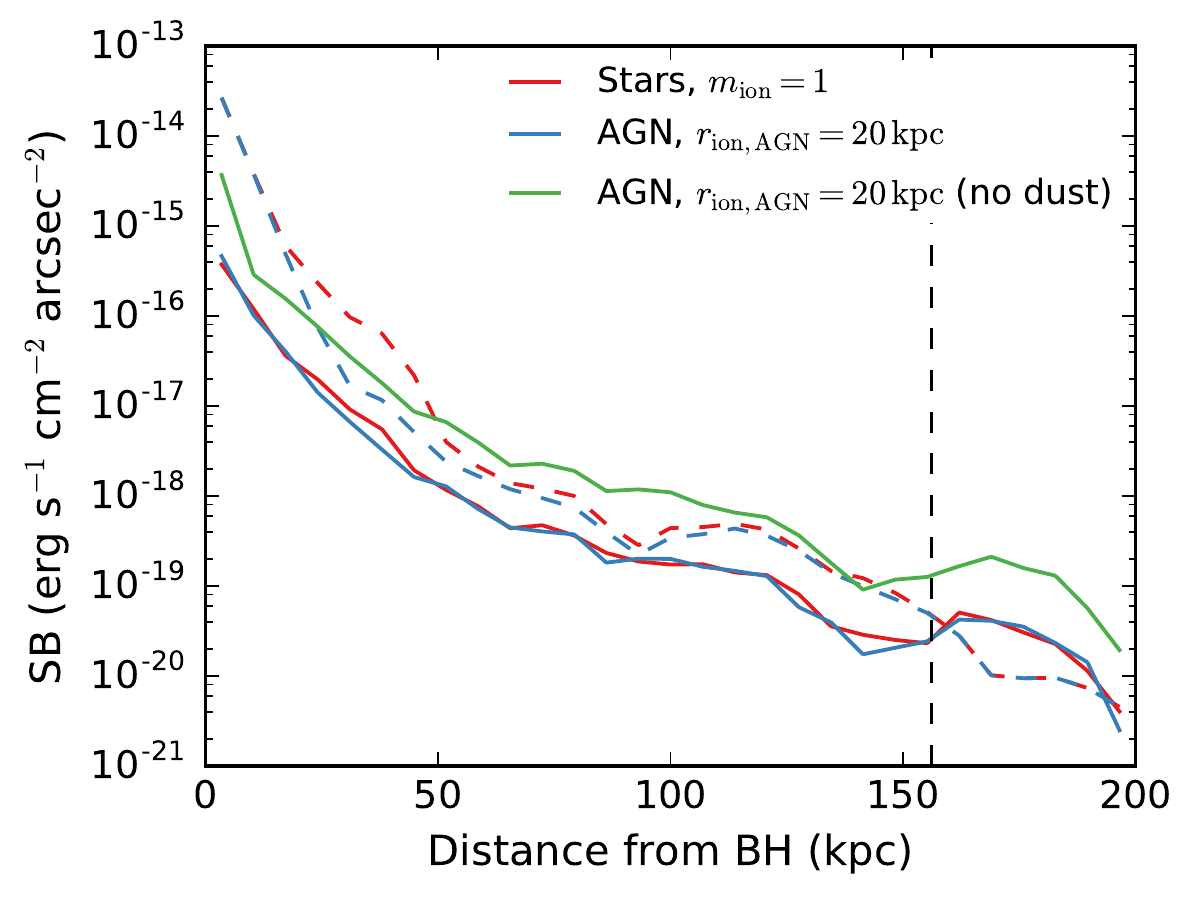}
  \caption{Radially binned surface brightness profile as a function of distance from the black hole for the fiducial models as `seen' by an observer directed in the same way as in Fig.~\ref{fig:SB}. The dashed and solid lines show the (intrinsic) surface brightness profile before and after the radiative transfer calculations, respectively. The vertical line denotes the virial radius of this halo ($R_{200c}$).}
  \label{fig:SB_profile}
\end{figure}

\subsection{Radiative transfer calculations}
\label{sec:radi-transf-calc}

Prior to carrying out the full \Lya radiative transfer calculation we interpolate the individual halo from the hydrodynamical simulation to a Cartesian grid with $512\times 528\times 449$ cells (corresponding to a uniform side-length of each cell of $\sim 1.63\,$kpc) while keeping constant the number of hydrogen atoms, the number of dust grains, the number of clouds placed on the grid and the total \Lya luminosity. For the conversion from the moving-mesh structure to a Cartesian grid (used by \texttt{AREPO} \citep{2010MNRAS.401..791S} and the radiative transfer calculations, respectively), we used \texttt{splash} \citep{2012JCoPh.231..759P} with a smoothing length\footnote{We used the default ${\rm M}_4$ cubic B-spline kernel.} of $l_{\rm smooth}= 4\times V_{\rm cell}^{1/3}$ which corresponds to $\sim 58$ neighboring particles. We repeated our analysis for one model using half the number of cells per dimension and find the results to be unaffected.
After the conversion the number of clouds in each cell is rounded to the nearest integer, redistributing the cutoff material self-consistently and the clouds are placed randomly within each cell.

The \Lya radiative transfer calculations are carried out using the Monte-Carlo radiative transfer (MCRT) code \texttt{tlac} \citep{Gronke2014a}. General descriptions of MCRT are given in \citet{Dijkstra2014_review} or \citet{Laursen2010}. The specific settings of \texttt{tlac} employed are identical to \citet{Gronke2016_clumps} and we merely summarize them here. We ran each MCRT simulation using at least $2\times 10^6$ photon packages which we placed randomly on the gas density grid. The probability to position a photon package in a certain cell (i.e., the weight of this cell) was proportional to the cell's intrinsic luminosity. We draw the intrinsic frequency of a photon package from a Voigt profile, the convolution of the natural (Lorentzian) line profile, and the thermal (Gaussian) profile of the emitting atoms of the initial cell. The process of image making with \texttt{tlac} has not been published previously and we describe the method below.

Producing SB maps with MCRT can be challenging as the number of individual photon packages escaping in a specific direction is essentially zero. Therefore, we use a commonly used technique sometimes called the `peeling' algorithm \citep[][]{1984ApJ...278..186Y,Zheng2002_dl,Laursen2010}. In every $N$th scatter event (including the emission step), the optical depth, $\tau_{\rm obs}(\nu_v)$, in the direction $\vek{k}_{\rm obs}$ is recorded. Here, $\nu_v$ is the frequency which the photon would have had if it had flown in the direction $\vek{k}_{\rm obs}$. This assigns a weight\footnote{Note, that although we consider the slightly differing redistribution functions for the scattering via the $2P_{1/2}$ and the $2P_{3/2}$ states for the full radiative transfer process we assume a uniform angular probability density function for the potential scattering towards $\vek{k}_{\rm obs}$. However, due to the large amount of photons that escaped without scattering ($\sim 35\%$) and thus, do not possess any preferred direction, we do not expect that our results are affected by this.} to this scattering event which is the escape probability along $\vek{k}_{\rm obs}$ given by:
\begin{equation}
w = \frac{N}{2} e^{-\tau_{\rm obs}(\nu_v)}\,,
\end{equation}
where $N$ is the frequency of recorded scattering events discussed above.
The SB for pixel $j$ can then be calculated using
\begin{equation}
SB_j = \frac{L_{\rm \alpha, i} S_j}{D_L^2(z) \Omega_{{\rm pix}, j} N_{\rm phot}}\,,
\label{eq:SB}
\end{equation}
where $N_{\rm phot}$ is the total number of photon packages emitted, $D_L(z)$ is the luminosity distance corresponding to redshift $z$, $\Omega_{{\rm pix}, j}$ is the solid angle of pixel $j$ and $S_j = \sum_j{w}$ is the sum of weights of the photons falling withing this pixel. Naturally, if calculating an observed spectrum, the weights $w$ have to be taken into account as well, and each frequency bin consists of the sum of the photons' weights.\\

\section{Results}
\label{sec:results}

In this section, we present first results from our `fiducial' halo (\S~\ref{sec:intr-lya-lumin}-~\ref{sec:lya-spectra}) and put them into a wider context in \S~\ref{sec:comp-observ}. Specifically, we carry out full \Lya radiative transfer simulations for the former part -- from which we show SB maps and profiles (\S~\ref{sec:intr-lya-lumin}) and \Lya spectra (\S~\ref{sec:lya-spectra}). In the latter part, we compare the full distribution of \Lya halos found in the \texttt{Illustris} simulations to observations.

\subsection{Intrinsic \Lya luminosities}
\label{sec:intr-lya-lumin}
Fig.~\ref{fig:Li} shows the intrinsic \Lya luminosity for the models presented in \S\,\ref{sec:model}. In each of the panels, the black dashed line shows the value if the whole grid (i.e., the selected halo) was ionized and the star symbol marks the fiducial values in each model. The two leftmost panels show that quite moderate ionization activity from either the star forming regions or the AGN is able to saturate the luminosity allowed by the gas distribution, as given by the hydrodynamical simulation output. Also noticeable is the proximity of the computed \Lya luminosity ($L_{\rm \alpha, i, max}\sim 2\times 10^{44}\ergps$) and the observed value by \citet{Cantalupo2014} ($\sim 2\times 10^{44}\ergps$ excluding the emission directly from the quasar). The luminosity saturates because the densest regions are already ionised. For the `stars' model this occurs at $m_{\rm ion}\gtrsim 0.3$  and for the `AGN' model for $r_{\rm ion}\gtrsim 15\,$kpc. For even greater radii the gas density drops off -- and so does the increase in luminosity.

\subsection{Surface brightness}
\label{sec:surf-brightn}
Fig.~\ref{fig:SB} shows the surface brightness maps for our two models (see \S\ref{sec:model}) and one specific observing direction computed as given by Eq.~\eqref{eq:SB}.
In order to resolve the SB maps sufficiently we ran the radiative transfer calculations with at least $2$ million photons, each leading to a minimum of $300,000$ recorded events for each of the four observers' directions. In Appendix~\ref{sec:intrinsic_sb_maps} we show the intrinsic SB maps, i.e., without radiative transfer as comparison.

Fig.~\ref{fig:SB} shows very similar topography in both the AGN and stars ionization models. The stars produce slightly more prominent features in the outskirts of the halo. Both cases have an maximal extent of $\sim 350\,{\rm kpc}$ for this particular sightline, a central SB of $\sim 10^{-15}\sbu$ which falls to $\sim 10^{-18}\,\sbu$ in the outer regions.  Overall, the size of this simulated nebula falls $\sim 100\,{\rm kpc}$ short of the projected maximum extent measured by \citet{Cantalupo2014}. However, if restricted to SB contours $\gtrsim 10^{-18}\sbu$ then the measurement is reduced by $\sim 20\%$ \citep{Cantalupo2014} which brings the two measures very close to each other. Fig.~\ref{fig:SB} also shows the `AGN' case but without the inclusion of dust (rightmost panel). Here, one can notice that the morphology and SB levels are different, that is, in the case without dust the halo is extended and overall brighter.

Fig.~\ref{fig:SB_profile} shows a more quantitative comparison of the surface brightness between the two models. Here we display the SB profiles, as well as the intrinsic SB profiles, as a function of distance from the black hole, observed from the same direction as in Fig.~\ref{fig:SB}. The same features are visible, with the curves of the `stars' and `AGN' cases following each other closely. Note, however, that the SB in this case is an angular averaged value and thus dependent on the geometry of the halo. This leads to variations in the radial profile depending on the observer's direction. The black dashed line in Fig.~\ref{fig:SB_profile} denotes the `virial radius' $R_{200c}$, i.e., the radius at which the average density is $200$ times the critical density at $z\sim 2$. In all cases the central SB values are significantly reduced compared to their intrinsic values due to dust extinction \citep[also see][]{Laursen2009} and re-distribution of photons due to radiative transfer effects.  However, in the outer parts of the halo the intrinsic and post-processed SB values approach each other.

In order to be able to distinguish between the effect of dust attenuation and systematic radial re-distribution of photons due to radiative transfer effects we also show in Fig.~\ref{fig:SB_profile} the `AGN' case without any dust (green line). Radiative transfer reduces the SB values in the inner part of the halo, but increases them at larger radii. Thus scattering increases the effective size of \Lya halos \citep[as also found by][in their dust-free simulations]{Trebitsch2016}. Note though that these re-distributed photons will also experience a relatively large dust optical depth, leading to lower SB values (as can be seen when comparing the two `AGN' curves in Fig.~\ref{fig:SB_profile}).
When considering, for example, the width of a halo with SB $> 10^{-18}\sbu$, these different effects would lead to uncertainties of a factor of $\sim 2$. In particular, hydrogen scattering tends to increase the apparent size 
compared to the intrinsic emission while dust absorption decreases it.
This may be of importance when comparing to observations and will be discussed again in \S~\ref{sec:comp-observ}.

\begin{figure}
  \centering
  \plotone{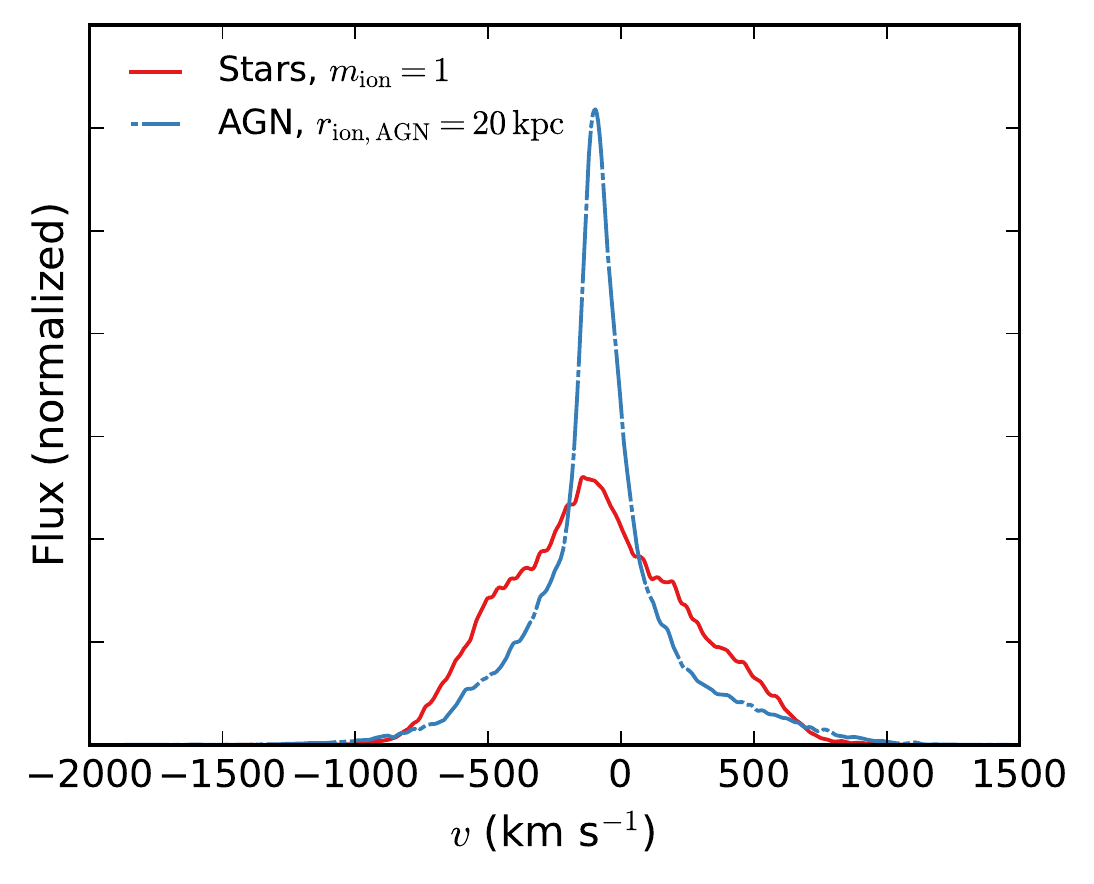}
  \caption{\Lya spectra taken from both our models in the direction used in Fig.~\ref{fig:SB_profile}. Note that the shapes of the spectra depend strongly on our fiducial model parameters (see \S\ref{sec:lya-spectra}).}
  \label{fig:spectra}
\end{figure}

\subsection{\Lya spectra}
\label{sec:lya-spectra}

Fig.~\ref{fig:spectra} shows spectra resulting from the full \Lya radiative transfer simulations. These spectra are taken from the same direction as the SB values from \S\,\ref{sec:surf-brightn}. The spectra are single peaked, showing the features expected from a low optical depth system.
While the `AGN' case shows a peaked profile, the spectrum emerging from our `stars' model is more rounded and wider, with a slightly peaked feature at the line center.

One has to be cautious, however, when interpreting the \Lya spectra shown in Fig.~\ref{fig:spectra}, as the applied resolution is not sufficient to capture the full \Lya radiative transfer dynamics \citep[see, e.g.,][who showed that $\sim$ pc resolution is necessary]{Verhamme2012}, and their shape heavily relies on the model parameters such as the minimally allowed neutral fraction for ionized cells $X_{\rm HI, min}$. This value -- which is zero for our fiducial models -- strongly influences the emergent spectral shape. We tested this by increasing $X_{\rm HI, min}$ to the rather extreme value of $10^{-3}$ in the `AGN' case and obtained wide (peak separation of $\sim 1500\kms$) double peaked spectra instead. In addition, we find the spectral shape to be dependent on the observer's direction. In particular, the flux at line center is significantly lower (however, not so low as to form a double-peaked profile) for other directions in which the optical depth between the observer and the emitting region is higher. This illustrates the difficulty using ab initio hydro-dynamical simulations to predict the outcome of \Lya spectra and their comparison with observations (see \S~\ref{sec:conclusions} for a discussion of this point).

\begin{figure}
  \centering
  \plotone{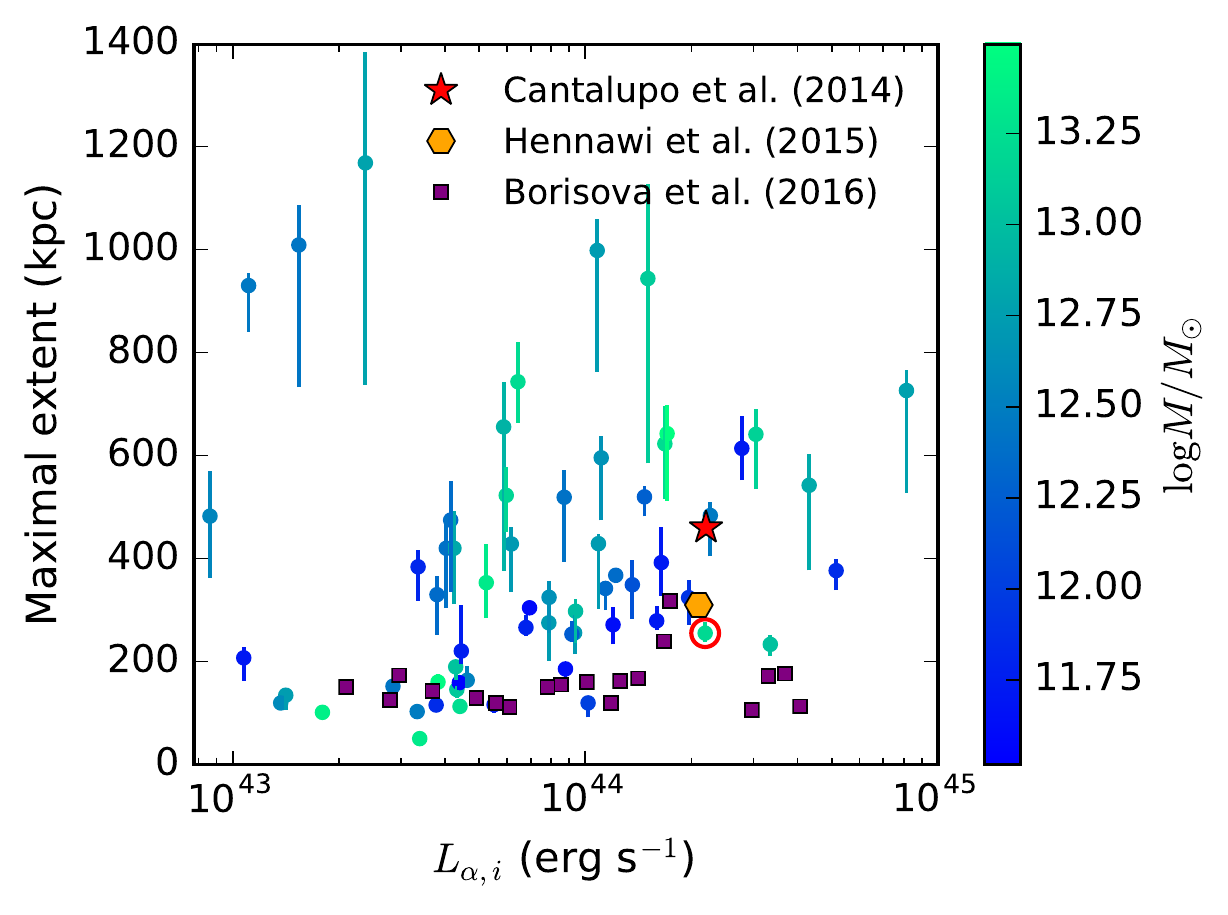}
  \caption{Maximal extent for halos in mass bins from $10^{11.5}\,M_\sun$ to $10^{13.5}\,M_\sun$ using a SB cutoff of $10^{-18}\sbu$ versus intrinsic \Lya luminosity. 
The error bars denote the variance between various directions and the red circle highlights the `fiducial' halo shown in the previous figures.
Also shown are the observations from \citet{Cantalupo2014}, \citet{Hennawi2015} and \citet{Borisova2016} which implicitly assume $f_{\rm \alpha, esc}\sim 1$. See \S\,\ref{sec:comp-observ} for a detailed discussion.\\}
  \label{fig:L_vs_extent}
\end{figure}

\subsection{Comparison to observations}
\label{sec:comp-observ}

\citet{Cantalupo2014} observed a giant \Lya halo with luminosity of $L_\alpha= (2.2\pm 0.2)\times 10^{44}\ergps$ ($L_\alpha \approx 1.43\times 10^{45}\ergps$ including the central quasar) and maximal projected extent of $460\,$kpc ($SB > 10^{-18}\sbu$)\footnote{We acknowledge that it is difficult observationally to distinguish between quasar and nebula emission. Note however the luminosity \textit{including} the quasar can be seen as an upper limit to the nebular emission.}. A similar -- slightly smaller -- object was observed by \citet{Hennawi2015} with a total \Lya luminosity of $L_\alpha \approx 2.1\times 10^{44}\ergps$ and a maximum extent of $310\,$kpc. 
The $\sim 10^{12.5}\,M_{\sun}$ halo we selected from the \texttt{Illustris} simulation shows very similar intrinsic \Lya luminosities ($L_{\rm i, \alpha} \sim 2\times 10^{44}\ergps$) and extents even without introducing additional clumping below the simulation resolution. In our fiducial models, the \Lya escape fraction was always $\sim 25\%$ leading to a slightly smaller \Lya luminosity than observed.
However, as the escape fraction is given by small-scale radiative transfer physics we see the obtained escape fraction as a lower limit to the real value. In particular two points are rather uncertain. First, the conversion between metallicity and \Lya dust optical depth as given by Eq.~\eqref{eq:dust} is calibrated to local values and it is unclear how reliable it is when applied to $z\sim 2$ systems. Furthermore, if the dust is not spread uniformly (below the radiative transfer resolution scale), ``dust-free channels'' might enhance the \Lya escape fraction. 

As already mentioned in \S~\ref{sec:surf-brightn} the non-unity escape fractions and radiative transfer effects will have an impact on the maximal extent of the \Lya halo, too. In particular, for our `fiducial' halo presented previously we found that while the intrinsic maximum extent (for which $SB > 10^{-18}\sbu$) is $\sim 250\,$kpc (for the AGN and star ionization cases), the post-processed maximum extent is $\sim 318\,$kpc and $\sim 415\,$kpc with and without dust, respectively\footnote{These values are for the `AGN' case ($r_{\mathrm{ion}} = 20\,$kpc) and the mean of four directions orthogonal to each other. A smoothing with $FWHM=1\,$arcsec has been applied.}. That \Lya scattering tends to increase the observed size of an object is frequently inferred for other objects; galaxies, for instance, show significantly larger \Lya halos compared to their UV counterparts \citep[e.g.,][]{2014ApJ...782....6H,Wisotzki2015}.
Other theoretical work seems to confirm this picture \citep[e.g., ][found that including radiative-transfer effects enlarges the LAH in their simulation by $\sim 10\%$]{Trebitsch2016}.
We, therefore, conclude that the re-distribution of photons in the outer regions tends to increase the maximum extent of the halo, and use as a conservative measure the intrinsic values in this section. We also note that uncertainties in the impact of \Lya radiative transfer persist as the \HI and dust structure on the smallest scales is unknown -- which might alter our results.

\subsubsection{Maximal extents}
In order to determine whether this halo is an exceptional case -- without running the full \Lya radiative transfer simulations on several halos -- we used the intrinsic \Lya luminosity as given by Eq.~\eqref{eq:Li} to compute `intrinsic surface brightness maps' from which we can measure the maximum projected extent. This approach is supported by the findings presented in \S~\ref{sec:surf-brightn}, namely that radiative transfer effects and dust extinction dim the central brightest regions of the halo, but only moderately increase the outer regions. Since we are interested primarily in these regions when comparing the simulations to observed size measurements we affect using intrinsic luminosities will only moderately
underestimate the maximal extents in our simulated halos.
Specifically, we used $15$ halos per $\log \Delta M / M_\sun = 0.5$ mass bin for the masses between $10^{11.5}M_\sun$ and $10^{13.5} M_\sun$ which we ionized according to our `ionization from stars' model\footnote{As shown in \S\,\ref{sec:results} the `AGN' and `star' case are very similar and also an increase of $m_{\rm ion}$ above $\sim 0.3$ does not alter $L_{\rm \alpha,i}$. Because not all (sub)halos possess black holes, however, we chose the `star' model out of simplicity.} with $m_{\rm ion}=1$. To mimic the observations, we then smooth the SB maps with a Gaussian kernel ($FWHM = 1\,$arcsec) and then measure the maximal extent for pixels with $SB_i > 10^{-18}\sbu$.

Fig.~\ref{fig:L_vs_extent} shows the result of this analysis as the maximal extent versus the intrinsic luminosities of the $45$ halos. The points and error bars represent the $16$th, $50$th and $84$th percentiles of the distribution of extents created though $100$ randomly drawn observer's directions, and the color coding denotes the mass of the halo. Note that the size of the error bars thus shows a substantial size-variation with viewing angle. In addition, we displayed the observations quoted above. These are not, however, the intrinsic values. $L_{\rm \alpha,i}$ for the observed nebula is probably a factor of a few larger, and $SB_{\rm i}$ should also be (slightly) increased. Nevertheless, Fig.~\ref{fig:L_vs_extent} illustrates that \Lya halos with similar extent \& luminosity can be found in the \texttt{Illustris} simulation. Another interesting feature is that the extent does not seem to correlate with mass and/or $L_{\rm \alpha, i}$. We also checked the correlation of the extent versus various other halo properties (gas mass, gas metallicity, black hole mass) and found none of them to correlate significantly with the extent (all the Pearson as well as the Spearman correlation coefficients were within $[-0.3,\,0.3]$).
The reason for this is that the quantity `maximal projected extent' is primarily a geometrical measurement and, hence, highly dependent on the topology of the gas configuration. 

\begin{figure}
  \centering
  \plotone{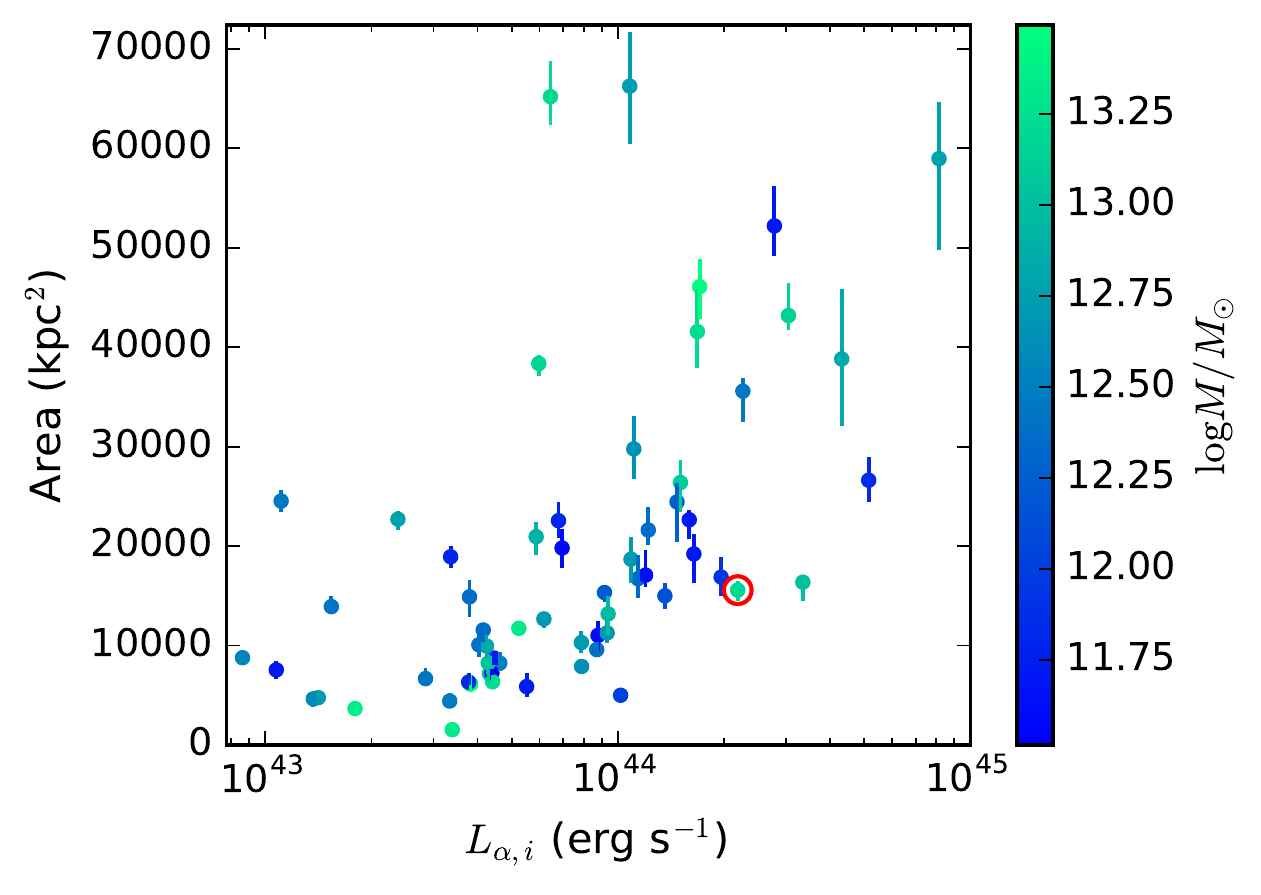}
  \caption{Projected area for halos in mass bins from $10^{11.5}\,M_\sun$ to $10^{13.5}\,M_\sun$ using a SB cutoff of $10^{-18}\sbu$ versus intrinsic \Lya luminosity. 
The error bars denote the variance between $100$ randomly drawn directions and the red circle marks the `fiducial' halo shown in Figs.~\ref{fig:Li}-\ref{fig:SB_profile}.\\}
  \label{fig:L_vs_extent_area}
\end{figure}

\begin{figure}
  \centering
  \plotone{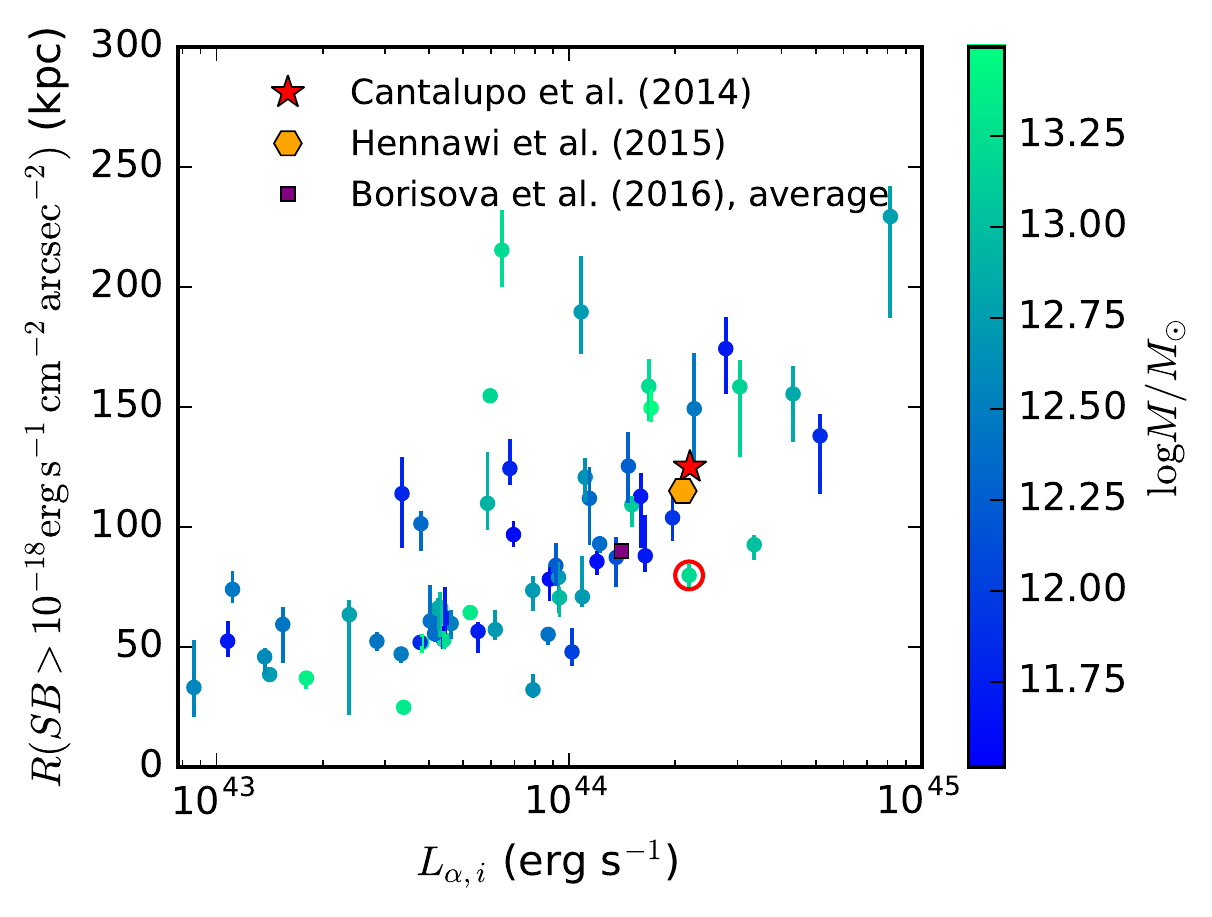}
  \caption{\Lya luminosity radius in which the SB (radially averaged) is $> 10^{-18}\,\sbu$ versus \Lya luminosity for halos in mass bins from $10^{11.5}\,M_\sun$ to $10^{13.5}\,M_\sun$ using a SB cutoff of $10^{-18}\sbu$.
As previously, the error bars denote the variance between $100$ randomly drawn directions and the red circle highlights our `fiducial halo'.
Also shown are the observations from \citet{Cantalupo2014} and \citet{Hennawi2015} \citep[estimated from Fig.~12 of][]{Battaia2016} as well as the average profile of \citet{Borisova2016} which implicitly assumes $f_{\rm \alpha, esc}\sim 1$ (see \S\,\ref{sec:comp-observ}).
\\}
  \label{fig:L_vs_extent_radial}
\end{figure}

\subsubsection{Alternative size measurements}
\label{sec:altern-size-meas}
As an alternative to the `maximal extent' measure, we plot in Fig.~\ref{fig:L_vs_extent_area} and Fig.~\ref{fig:L_vs_extent_radial} the projected area covered by the LAHs, and the radius for which the radially averaged SB profile falls below $SB_{\rm cut}$, respectively. Again we used a cutoff of $SB_{\rm cut} = 10^{-18}\sbu$ and a FWHM of the convolved Gaussian kernel of $1\,$arcsec. It is noticeable that the correlation between area covered as well as the `crossing radius' and intrinsic \Lya luminosity is slightly better than between the maximal extent and $L_{\rm \alpha,i}$ shown in Fig.~\ref{fig:L_vs_extent} (Spearman coefficients of $\sim 0.65$ and $\sim 0.72$ versus $\sim 0.34$). However, a significant scatter still persists.
Fig.~\ref{fig:L_vs_extent_radial} shows as a comparison the objects from \citet{Cantalupo2014}, \citet{Hennawi2015} and \citet{Borisova2016}. The radial extent of the former two was estimated from Fig.~12 of \citet{Battaia2016}, and thus implicitly assumes a \Lya fraction of unity -- as already discussed above for Fig.~\ref{fig:L_vs_extent}. 

\begin{figure}
  \centering
  \plotone{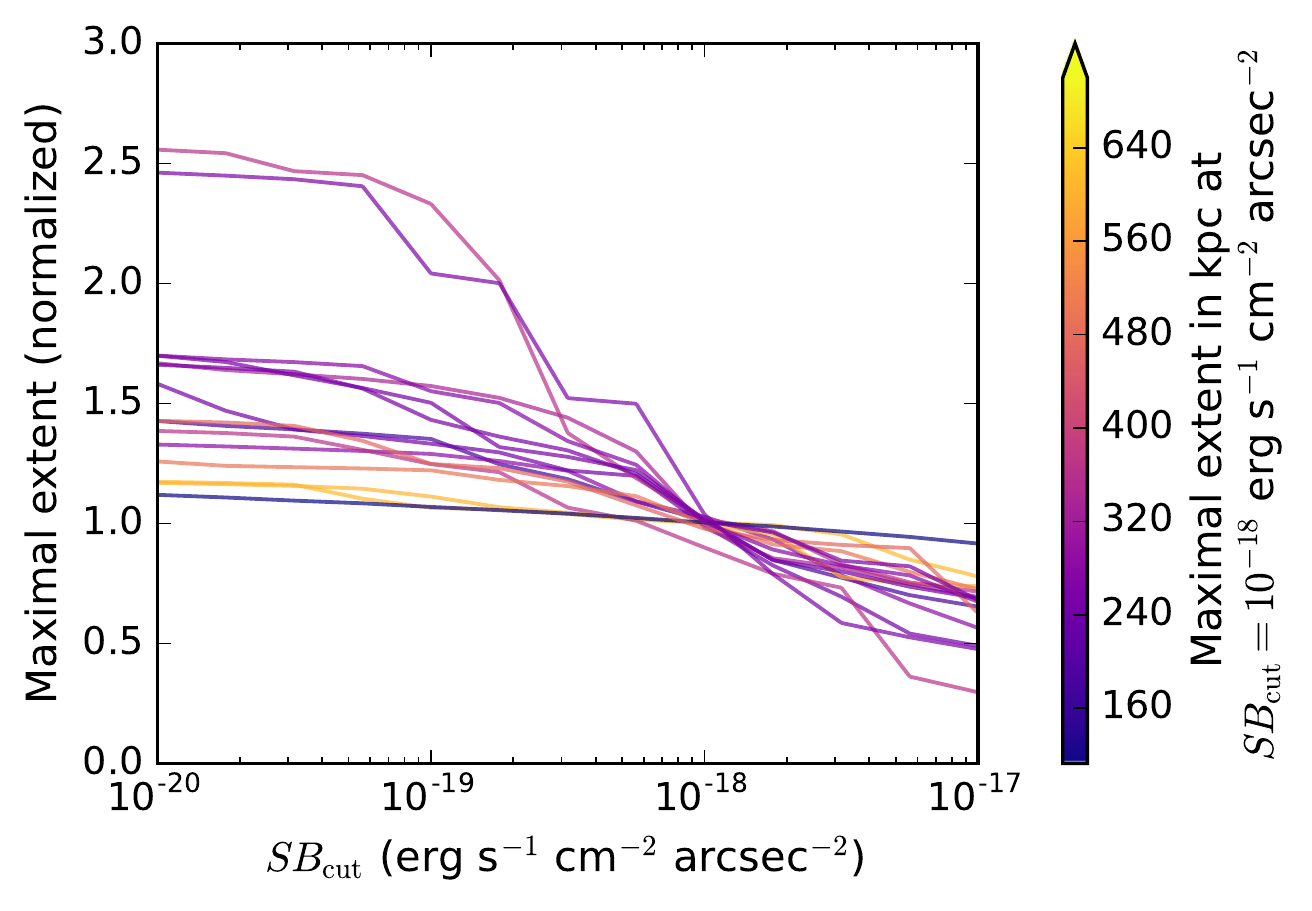}
  \caption{Maximal extent versus the surface brightness cutoff $SB_{\rm cut}$. Each line represents one of $15$ randomly selected halos for which we plot the mean (using $10$ random observer directions) normalized to the value at $SB_{\rm cut}=10^{-18}\sbu$ (for $100$ random observer directions). The color coding denotes the value of normalization. 
}
  \label{fig:extent_vs_sbcut}
\end{figure}

\subsubsection{Size dependence on SB cutoff}
\label{sec:size-dependence-sb}
It initially appears puzzling that \Lya halos with this extent have not been detected previously, as is nicely illustrated in Fig.~3 of \citet{Cantalupo2014}. However, as noted in that work, one has to take into account that previous halos were measured with a surface cutoff of $SB_{\rm cut}=3\times 10^{-18}\sbu$ whereas \citet{Cantalupo2014} used $SB_{\rm cut}=10^{-18}\sbu$. That this difference matters is shown in Fig.~\ref{fig:extent_vs_sbcut} where we plot how the maximal extent varies with chosen surface brightness cutoff $SB_{\rm cut}$. In particular, we show\footnote{For illustration purposes we show only $15$ randomly selected halos.} the mean of the maximal extent using $10$ random observer directions and normalized it to the mean of $100$ random observer directions for $SB_{\rm cut} = 10^{-18}\sbu$ (as used in Fig.~\ref{fig:L_vs_extent}). Fig.~\ref{fig:extent_vs_sbcut} shows that the exact value of $SB_{\rm cut}$ can alter the measured projected extent dramatically and even a moderate change of factor three as discussed above can decrease the size by $\sim 50\%$. Interestingly, lowering $SB_{\rm cut}$  to $10^{-19}\sbu$ might more than double the size but even fainter cutoffs do not increase the extent further. 

While we were finalizing this work, two new studies appeared examining extended LAHs. \citet{Battaia2016} examined $15$ quasars at $z\sim 2$ and found no extended nebulae. By contrast, \citet{Borisova2016} examined a sample of $17$ radio-quiet quasars at $3 < z < 4$ and found a nebula with extent $>100\,{\rm kpc}$ in every case examined \citep[see also][who found in $4$ out of $5$ cases no extended nebula]{Herenz2015a}. There thus seems at present some disagreement as to the abundance of these objects. Whether this is due to the differing samples and redshift ranges, or, as discussed in \citet{Borisova2016}, differing observational techniques, is beyond the scope of this work. However, we note that our simulations predict fairly ubiquitous LAHs in the presence of even moderate ionizing flux from the central black hole; in our simulated sample of quasars, only $\sim 15\%$ had nebular emission below the observable threshold of $10^{-18}\sbu <50$ kpc from the halo (radially averaged). The predicted detection rate would thus appear to be in good agreement with the results of \citet{Borisova2016}. Confirmation of a substantially lower detection rate would thus suggest that not all active quasars contain the ionizing photons necessary in our model to produce nebulae. This could be realised either if quasar emission were tightly beamed or if the quasar lifetime were short. 

\subsubsection{Surface brightness profiles}
\label{sec:surf-brightn-prof}
\begin{figure}
  \centering
  \plotone{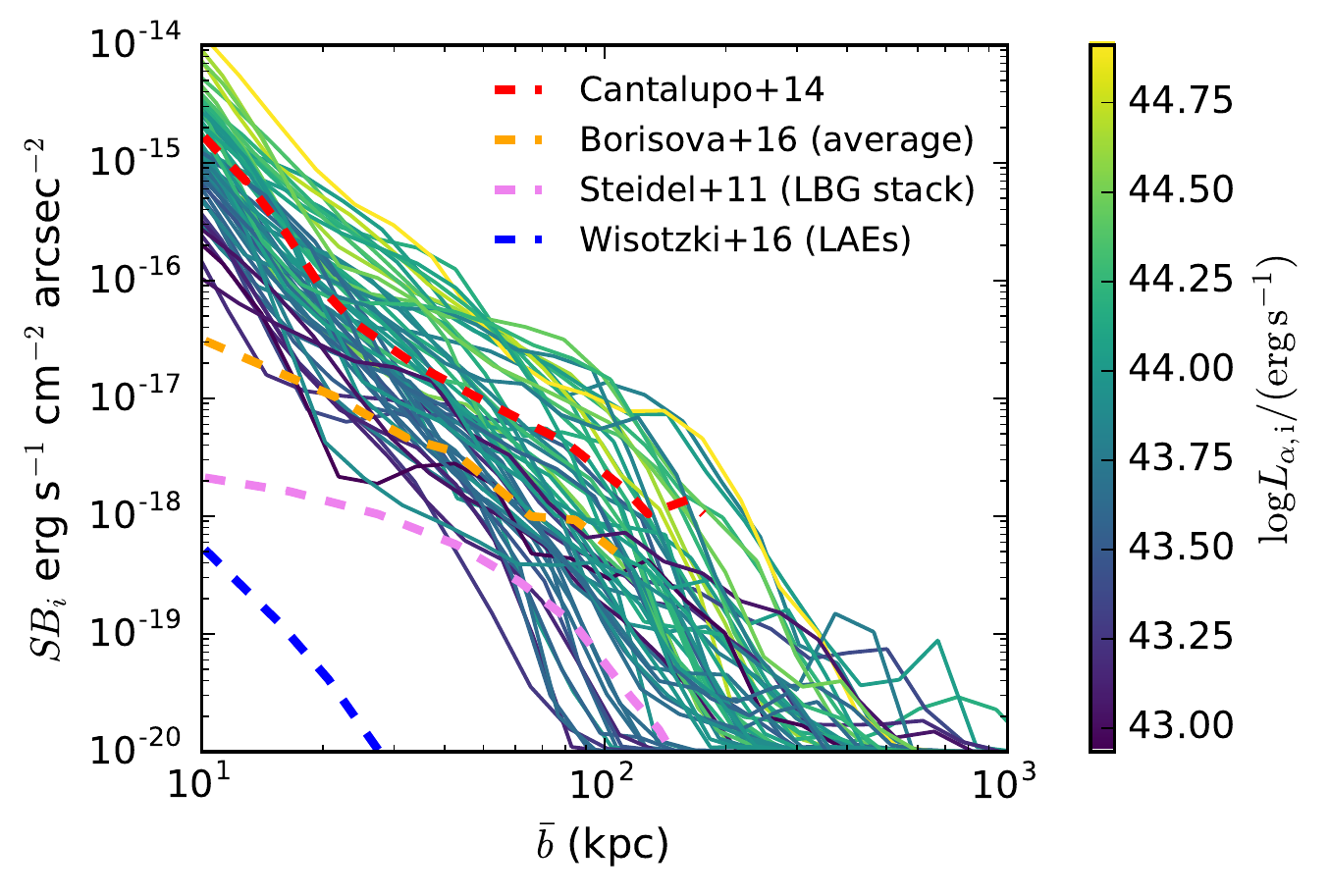}
  \caption{Circularly averaged intrinsic surface brightness profile for our sample of $60$ nebulae between $10^{11.5}M_\sun$ and $10^{13.5}M_\sun$. Each solid curve represents the average $SB_{\rm i}$ profiles assembled from $10$ randomly drawn viewing angles and is color coded according to the intrinsic \Lya luminosity. The dashed lines show the data from \citet{Cantalupo2014}, \citet{Borisova2016}, \citet{Steidel2011} and \citet{Wisotzki2015} (from top to bottom, extracted from figure 5 of \citet{Borisova2016}) as comparison. Note, however, that the data shows naturally the measured  $SB$ values and not the intrinsic ones (see \S~\ref{sec:comp-observ} for a discussion).
}
  \label{fig:SBi_profiles}
\end{figure}

In Fig.~\ref{fig:SBi_profiles} we compare the full circularly averaged surface brightness profiles to the results of \citet{Cantalupo2014} and \citet{Borisova2016}. Again, we used the `stars' model with $m_{\rm ion} = 1$ and a smoothing kernel with FWHM $ = 1\,$arcsec. Overall the agreement is very good, however, note that our profiles (solid lines) show the intrinsic SB values wheres the observations (dashed lines) measure the SB after radiative transfer effects. This means our curves represent an upper limit on the measured SB levels due to the destruction of \Lya photons by dust.  In particular, comparing the curve of \citet{Cantalupo2014} to our findings suggest an overall homogeneous escape fraction of $\gtrsim 10\%$ as the slopes are comparable. The (averaged) SB profile of \citet{Borisova2016} is flatter in the inner region which could be due to a higher dust attenuation the central region and re-distribution to larger radii due to scatterings. This is not unreasonable as the central region possesses a larger optical depth which makes photons originating there more vulnerable to destruction by dust and more likely to diffuse significantly in space before escape. 
We do in fact find this in the full radiative transfer simulations in \S\ref{sec:surf-brightn} (comparing the intrinsic and post-processed SB profiles in Fig.~\ref{fig:SB_profile}).

\section{Conclusions}
\label{sec:conclusions}
Several \Lya halos have recently been detected with unusually large extent ($\gtrsim 300\,$kpc) and observed \Lya luminosity ($\gtrsim 10^{44}\ergps$) \citep{Cantalupo2014,Hennawi2015}. It has been speculated that this suggests the presence of substantial amounts of cold gas in cool clumps at densities which are not resolved by modern hydro-dynamical simulations.

Using the publicly available data of the \texttt{Illustris} simulation \citep{2014Natur.509..177V,2015A&C....13...12N} we modelled the \Lya emission emerging from large halos ($M > 10^{11.5}M_{\sun}$) at $z\sim 2$. We considered two simple models: an AGN as source of ionizing photons and ionization due to stars. We found for a single halo where we performed full \Lya radiative transfer that \textit{(i)} with a moderate strong ionization source the halo showed intrinsic \Lya luminosities and extents comparable to observational data; \textit{(ii)} additional clumping does not seem necessary to explain the first order properties of the observed giant LAHs, and, \textit{(iii)} due to the low optical depth of some routes escaping \Lya photons do not scatter many times leading to a single peaked \Lya profile in both models. 
While the difference in the SB profiles cannot be used to distinguish between the models, the \Lya spectra are sensitive to sub-resolution properties such as the kinematics of the ISM and, thus, the question whether or not the \Lya spectrum contains information about the main ionization source is still outstanding. In particular, changing the ionization state in the maximally ionized regions slightly leads to a much greater optical depth at line center and, thus, an emergent double peaked as opposed to the single peaked profiles in our fiducial models. This difficulty might be the cause between observed spectra of \Lya nebulae which are often double-peaked with an extended red- or blue-tail
\citep[e.g.][]{2006ApJ...640L.123M,2016arXiv160703112V} and cosmological hydro-dynamical simulations which often predict a single peaked profile -- as in this work \citep[also see, e.g.,][]{Trebitsch2016}. 

In order to put our fiducial halo into a wider context, we computed the intrinsic \Lya SB maps
of $45$ halos in the mass range $11.5 < \log M / M_{\sun} < 13.5$ and find that the observed objects fall within the range of computed maximal projected extents and \Lya luminosities. Furthermore, we find that that neither the maximal extent nor the total projected area of \Lya halos correlates significantly with other halo properties such as the total (intrinsic) \Lya luminosity which we attribute to the stochastic nature of the gas-morphology.

Varying the surface brightness cutoff used for characterizing the extent of the \Lya halos, we find that multiplying (dividing) this value by factor of three can increase (decrease) the maximal extent by $\sim 50\%$. This means not only that the nature of the previously detected `giant' \Lya halos might not be that different to the smaller ones detected previously but also that we expect to find more of these objects in the near future.

This conclusion -- i.e., that `giant LAHs' do appear in modern hydro-dynamical simulations -- differs from \citet{Cantalupo2014} who performed a similar analysis on one $\sim 10^{12.5}\,M_\odot$ halo extracted from a hydro-dynamical simulation and found that substantial extra small-scale gas clumping was necessary to match their observations.
We think this is for two reasons: \textit{(i)} although their simulation included supernovae feedback it did not include AGN feedback which leads to a non-negligible distribution of gas for halos in this mass range \citep[see, e.g.,][for an illustration of this effect]{2014MNRAS.440.2997V,2014MNRAS.445..175G}; \textit{(ii)} as we show in Fig.~\ref{fig:L_vs_extent} there is significant scatter in \Lya extents -- even for fixed luminosity and / or size. Thus, a single halo producing a \Lya SB morphology as computed by \citet{Cantalupo2014} is entirely possible, even within Illustris.

\acknowledgments
The authors thank the anonymous referee for the constructive comments that significantly improved the manuscript. 
We thank M. Dijkstra and Ll. Mas-Ribas for a critical read of the draft.
MG thanks the Physics \& Astronomy department of Johns Hopkins University for their kind hospitality. SB was supported by NASA through Einstein Postdoctoral Fellowship Award Number PF5-160133.
This research made use of Astropy, a community-developed core Python package for Astronomy \citep{2013A&A...558A..33A}; matplotlib, a Python library for publication quality graphics \citep{Hunter:2007}; SciPy \citep{jones_scipy_2001}.

\bibliography{references_all}

\appendix

\section{Intrinsic SB maps}
\label{sec:intrinsic_sb_maps}
\begin{figure*}
  \centering
  \plotone{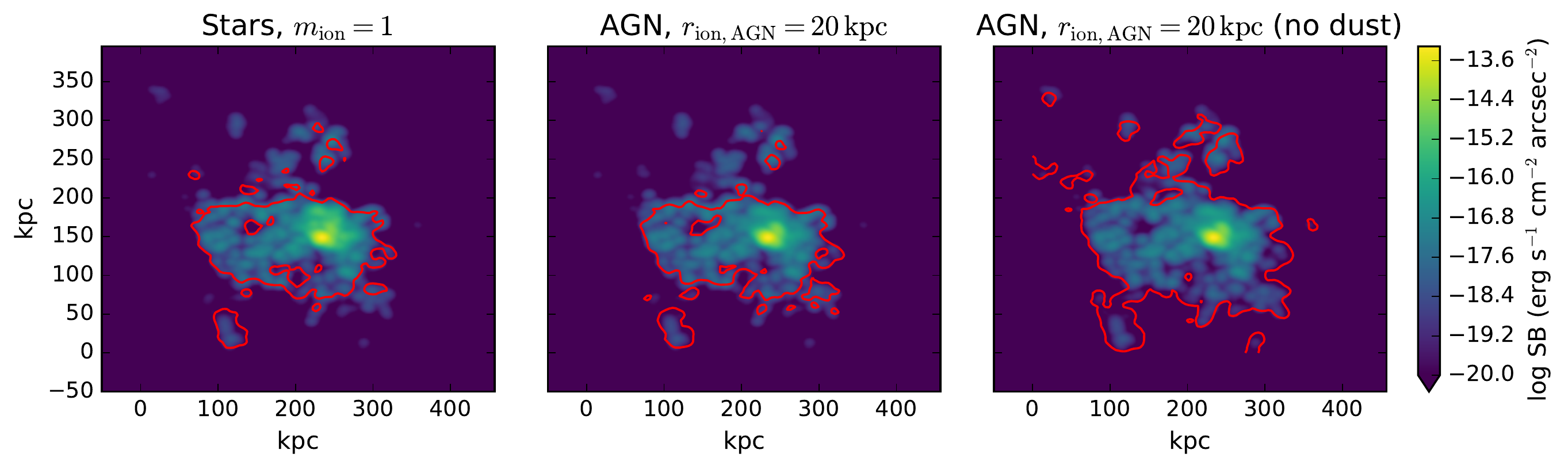}
  \caption{Intrinsic surface brightness maps for the fiducial models (see \S\ref{sec:model}). The red lines denotes the $SB = 10^{-18}\sbu$ contours from of the full radiative transfer simulation (see \S~\ref{sec:surf-brightn} and Fig.~\ref{fig:SB})\\}
  \label{fig:SB_intrinsic}
\end{figure*}

Fig.~\ref{fig:SB_intrinsic} shows the intrinsic (i.e. without radiative transfer effects and (dust) extinction) surface brightness maps. These surface brightness maps have been assembled as described in \S~\ref{sec:comp-observ} using a smoothing kernel with FWHM=$1\,$arcsec and from the same direction as Fig.~\ref{fig:SB}. In addition, we show in Fig.~\ref{fig:SB_intrinsic} the $SB = 10^{-18}\sbu$ levels from of the full radiative transfer simulation (\S~\ref{sec:surf-brightn}) also smoothed with the same kernel.

One can note that compared to Fig.~\ref{fig:SB} the intrinsic SB maps are firstly brighter in the central region due to the effect of dust extinction. This effect has been discussed in Sec.~\ref{sec:surf-brightn} (see Fig.~\ref{fig:SB_profile}). Secondly, the scattering enlarges and washes out the SB contours. However, the extent for $SB\gtrsim 10^{-18}\sbu$ is comparable.

\end{document}